\documentclass{jpsj-suppl}
\usepackage{txfonts} 

\usepackage{bm}

\newcommand{\Psfig}[2]{\includegraphics[width=#1]{#2}}
\newcommand{\Expect}[1]{\langle #1 \rangle}
\newcommand{\SUN}[1]{\text{SU} ( #1 )}
\newcommand{\UN}[1]{\text{U} ( #1 )}

\title{Quantum Fluctuations of Particles and Fields in Smooth Path Integrals}

\author{Takayasu \textsc{Sekihara}$^{1}$}

\inst{$^{1}$KEK Theory Center,
  Institute of Particle and Nuclear Studies, 
  High Energy Accelerator Research Organization (KEK),
  1-1, Oho, Tsukuba, Ibaraki, 305-0801, Japan}

\email{sekihara@post.kek.jp}

\recdate{July 14, 2013}

\abst{An approach to evaluation of the smooth Feynman path integrals
  is developed for the study of quantum fluctuations of particles and
  fields in Euclidean time-space.  The paths are described by sum of
  Gauss functions and are weighted with $\exp (-S)$ by appropriate
  methods.  The weighted smooth paths reproduce properties of the
  ground state of the harmonic oscillator in one dimension with high
  accuracy.  Quantum fluctuations of U(1) and SU(2) gauge fields in
  four dimensions are also evaluated in our approach.}

\kword{Path integrals, 
  quantum mechanics, 
  quantum field theory}

\begin{document}
\maketitle

\section{Introduction}

To clarify properties of the ground state of the system is one of the
most important problems in quantum physics.  For instance, the ground
state of the one-dimensional harmonic oscillator fluctuates around the
minimum of the potential due to the quantum effects in contrast to the
classical situation, in which the oscillator keeps to stay at the
minimum of the potential.  We can ``visualize'' such quantum
fluctuations by using the path integral method developed by
Feynman~\cite{Feynman:1948ur}, in which all possible paths are taken
into account with the probability amplitude $\exp (- S)$ in Euclidean
time-space, with $S$ the action of the system.

Although we cannot solve exactly the path integrals in almost all
cases, the evaluation of the path integrals can be simplified by
discretizing time-space, in which derivations and integrations are
replaced with finite differences and summations, respectively, and
measure of the path integrals becomes a countable product.  This
discretization, together with the weight $\exp (-S)$, allows us to
describe easily the ground state of the system both in numerical and
analytical calculations.  An important example of the discretized path
integrals is the lattice QCD (quantum
chromodynamics)~\cite{Wilson:1974sk}, by which nonperturbative aspects
of QCD have been revealed~\cite{Rothe:2005LGT}.

However, the time-space discretization also has some shortcomings.
Obviously, the time-space discretization explicitly breaks continuous
symmetries of time-space such as the translational symmetry down to
discrete symmetries.  Furthermore, the discretization sometimes leads
to qualitative discrepancies compared to the continuous theory such as
the famous doubler for the fermions on the
lattice~\cite{Wilson:1975QS} and magnetic monopoles in the lattice QED
(quantum electrodynamics)~\cite{Polyakov:1975rs}.  Since they are
brought by the time-space discretization, it is desired to perform the
path integrals in smooth time-space from the viewpoint of
complementarity for the discretized approach.

Motivated by these observations, we develop an approach to evaluating
the smooth path integrals in Euclidean time for particles and
fields~\cite{Sekihara:2012ng}.  The smooth paths are described by sum
of Gauss functions with weight $\exp (-S)$ by appropriate methods such
as the Metropolis method~\cite{Metropolis:1953am} and the heat-bath
method~\cite{Creutz:1980zw}.

\section{Basic ideas}

Firstly, for a nonrelativistic particle with one degree of freedom $q$
in a periodic boundary condition with period ${\cal T}$ in time
direction, $q(\tau + {\cal T}) = q(\tau )$, the path integral method
evaluates the quantum transition amplitude in Euclidean time
as~\cite{Feynman:1948ur},
\begin{equation}
{\cal Z} = \int _{\rm period} {\cal D} q \exp ( - S [ q ] ) ,  
\quad 
{\cal D} q \equiv 
\prod _{\tau} d q (\tau ) .
\label{eq:PI-Euclid}
\end{equation}
Since the quantum fluctuations of the particle are weighted
with the factor $\exp (-S)$, an expectation value of an operator
${\cal O}[q]$ in quantum mechanics can be evaluated by using $N$ paths
$q_{n}$ ($n=1,\, 2, \, \cdots , \, N$) which are 
appropriately weighted with $\exp (-S)$ as,
\begin{equation}
\Expect{{\cal O} [q]} 
= \frac{1}{\cal Z} \int _{\rm period} 
{\cal D} q {\cal O} [q] \exp ( - S )
\approx 
\frac{1}{N} \sum _{n=1}^{N} {\cal O} [ q_{n} ] , 
\end{equation}
where the last approximation becomes good for large $N$.  

Then, we consider the discretized approach to the path integrals for a
nonrelativistic particle in the periodic boundary condition.  In this
case the particle position is represented as $\tilde{q}_{i}$ at time
$\tilde{\tau} _{i} = i a$ with $i=1$, $\cdots$, $N_{\rm lat}$ and the
lattice spacing $a \equiv {\cal T} / N_{\rm lat}$, and the path of the
particle is obtained by connecting $\tilde{q}_{i}$ and
$\tilde{q}_{i+1}$ from $i=1$ to $N_{\rm lat}$ with straight lines.
Then measure of the path integral is now defined as a countable
product and the action is replaced with the corresponding discretized
one, $\tilde{S} [\tilde{q}]$.  The details of the discretized path
integrals for nonrelativistic particles are discussed in
Ref.~\cite{Creutz:1980gp}.

Now let us make our procedure for the simulation of the smooth path
integrals from analogy to the discretized approach.  Our idea is to
connect the neighboring points for the particle positions by smooth
lines rather than straight lines.  In order to achieve this, we smear
the particle position in the discretized approach $\tilde{q}_{i}$ with
a Gauss function of width $\xi _{i}$ at time $\tau _{i}$,
\begin{equation}
  \tilde{q}_{i} ~ (\text{at } \tau = \tilde{\tau} _{i}) ~~ \to ~~
  q_{i} \exp \left [ - \frac{(\tau - \tau _{i})^{2}}
    {\xi _{i}^{2}} \right ] ,
\end{equation}
for $i=1$ to $N_{\rm lat}$, where $\tau - \tau _{i}$ means to take
time distance between $\tau$ and $\tau _{i}$ in the periodic boundary
condition.  Here ($q_{i}$, $\tau _{i}$, $\xi_{i}$) in our approach
corresponds to ($\tilde{q}_{i}$, $\tilde{\tau} _{i}$, $a$) in the
discretized approach.  With this smearing one can naturally connect
the path with smooth lines rather than straight lines as,
\begin{equation}
  q (\tau ) 
  = \sum _{i=1}^{N_{\rm sum}} q_{i} 
  \exp \left [ - \frac{(\tau - \tau _{i})^{2}}
    {\xi _{i}^{2}} \right ] , 
  \label{eq:weighted-path}
\end{equation}
where $N_{\rm sum} (=N_{\rm lat})$ is number of the summed terms.  In
this construction of the smooth path the minimal scale of the
fluctuation for $q(\tau)$ corresponds to $\xi _{i}$.  The positions of
the Gauss functions $\tau _{i}$ may take random values rather than
values in same interval, $\tau _{i}= i {\cal T} / N_{\rm sum}$, as
long as every time is equally treated without making any special time.
The width of the Gauss function, or the scaling constant, $\xi _{i}$,
should be determined so as to prevent any special time coordinate.  In
this study we use a properly fixed value for $\xi$.  Then the weight
$\exp (-S)$ is given by appropriate methods for $q_{i}$ as in the
discretized approach~\cite{Creutz:1980gp}.
If the action $S[q]$ is expressed as polynomials of $q(\tau )$ and
$\dot{q} (\tau ) \equiv dq / d\tau$, the action can be rewritten as
polynomials of $q_{i}$ due to the expression of $q(\tau)$ in
Eq.~\eqref{eq:weighted-path} and properties of the Gauss functions,
and hence we can perform the heat-bath method~\cite{Creutz:1980zw} for
the weight $\exp [ - S (q_{i})]$ with respect to $q_{i}$.  If the
action $S[q]$ is not expressed as polynomials of $q_{i}$, one may use
the Metropolis method~\cite{Metropolis:1953am} by considering the
charge $\delta q_{i}$ with respect to $q_{i}$ and the corresponding
change of the action $S(q_{i})-S(q_{i}+\delta q_{i})$.  The update of
$q_{i}$ is performed until the action converges.

Here we note that $q (\tau)$ in Eq.~\eqref{eq:weighted-path} cannot
describe all possible paths, since the Gauss functions in
Eq.~\eqref{eq:weighted-path} cannot be a complete set with respect to
the smooth functions in the periodic boundary condition.  Therefore,
at this point our construction of paths is an approximation with
respect to the complete paths required by the measure ${\cal D}q$.
Nevertheless, we expect that most of the possible paths can be
effectively taken into account when value of $N_{\rm sum}$ is
sufficiently large with dense $\tau _{i}$ and small $\xi$.  We also
note that time uniformity may be broken when the time components $\tau
_{i}$ take values in same interval, $\tau _{i}= i {\cal T} / N_{\rm
  sum}$, but we expect that the uniformity will restore if one
considers sufficiently dense distribution of the Gauss functions.
Generalization to the systems with $f$ degrees of freedom ($f\ge 2$)
as well as to quantum field theories is obvious.

\section{Numerical results}

Now let us examine our approach by investigating a harmonic oscillator
in one dimension, which action is written as,
\begin{equation}
S_{\rm HO} = \int _{0}^{\cal T} d \tau L_{\rm HO} (q , \, \dot{q} ) , 
\quad 
L_{\rm HO} = \frac{1}{2} m \dot{q}^{2}  
+ \frac{1}{2} m \omega ^{2} q^{2} .
\label{eq:HOaction}
\end{equation}
Here we fix its mass and angular frequency as $m = \omega = 1$ and
take the time range ${\cal T} =20$.  We consider four simulation
conditions; $\xi = 2.0$ (A), $\xi=1.0$ (B), $\xi=0.5$ (C), and
$\xi=0.1$ (D), with the same number of the summation $N_{\rm sum}=200$
in each condition. For the initial condition we fix $\tau _{i}$ as
$\tau _{i}= i {\cal T} / N_{\rm sum}$ and randomly generate $q_{i}$ as
a hot start.  We prepare $N=400$ paths for each condition.  Since
temperature of the system $1/{\cal T}$ is much smaller than the
excitation energy, the quantum fluctuations in this study will reflect
the ground state of the harmonic oscillator.

\begin{figure}[b]
  \centering
  \begin{tabular}{c}
    \Psfig{7.50cm}{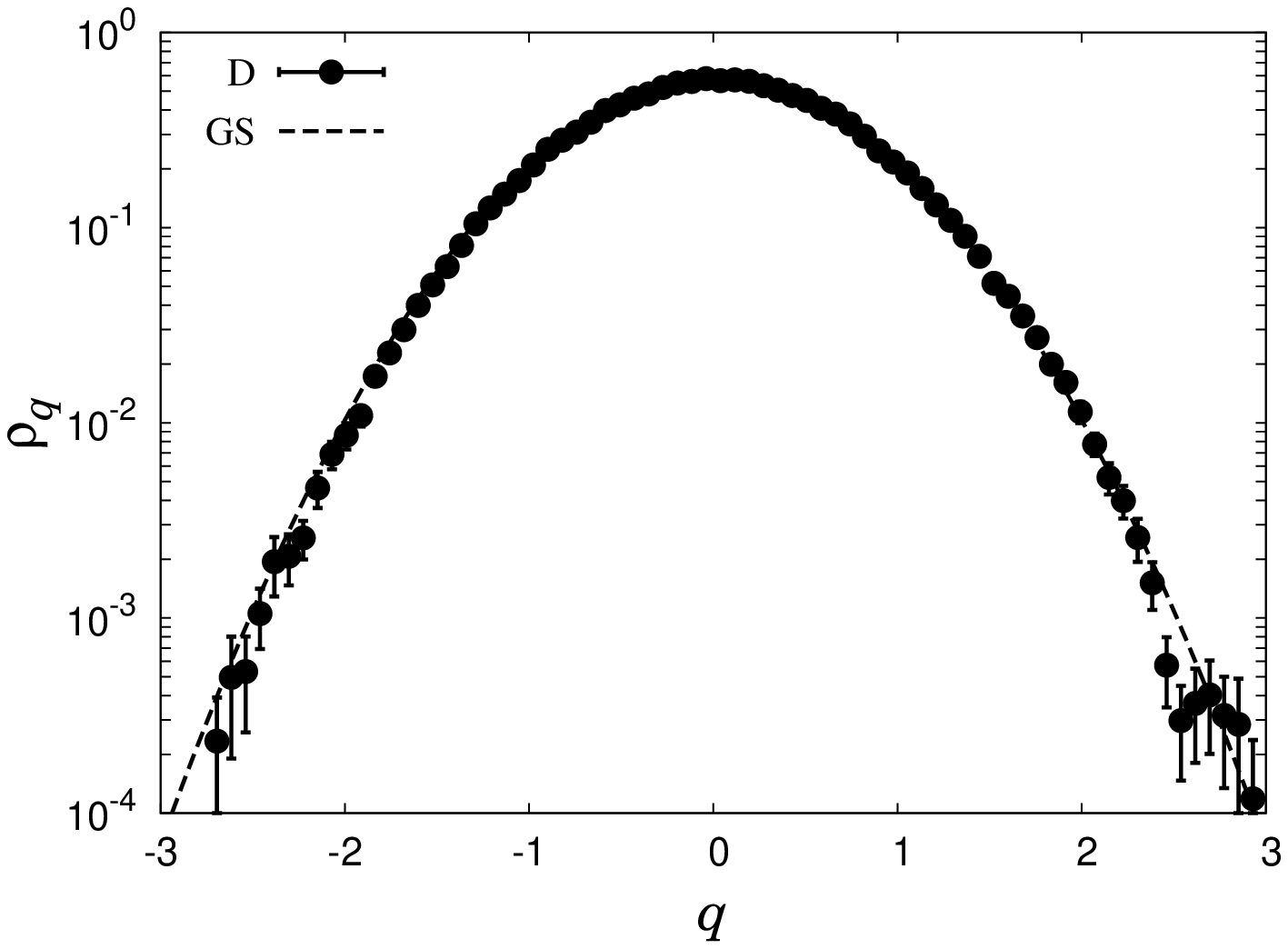}~\Psfig{7.50cm}{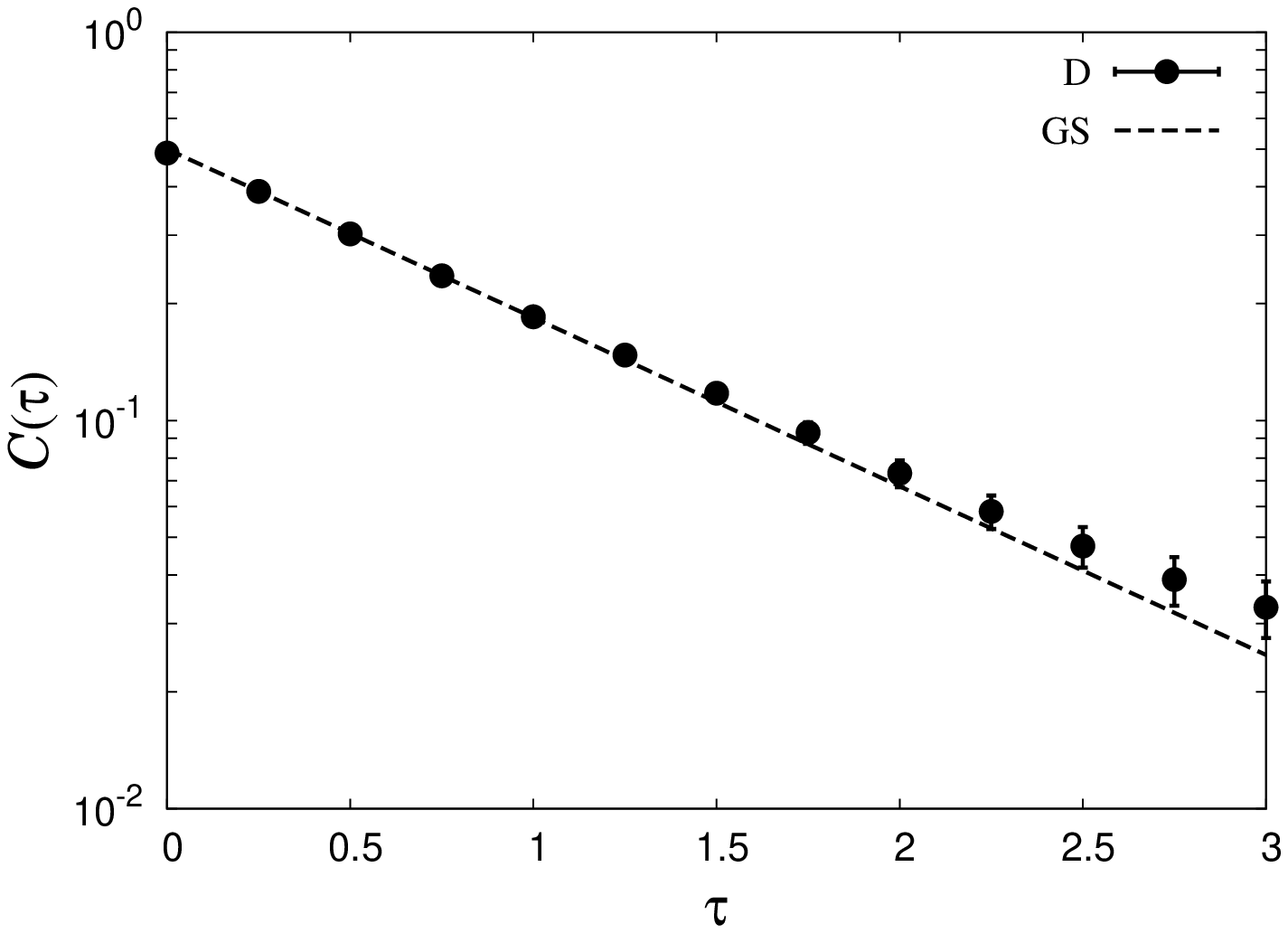} 
  \end{tabular}
  \caption{The distribution of the coordinate for the harmonic
    oscillator in condition D ($\xi = 0.1$ and $N_{\rm sum}=200$)
    together with the squared wave function of the ground state
    denoted by dashed line (left) and the correlation function for the
    fluctuation in condition D together with the exact value for the
    ground state denoted by dashed line (right).  Both are given in
    logarithmic scale.}
  \label{fig:HO}
\end{figure}

As a result of the numerical simulation, quantum paths approach to
equilibrium at around the iteration number $N_{\rm iteration}\sim 5
\times 10^{4}$ for all conditions.  After iteration $N_{\rm
  iteration}=5 \times 10^{4}$ the expectation value of the mean
squared radius of the fluctuation becomes $\Expect{\overline{q^{2}}}=
0.389 \pm 0.08$ for the condition A, $0.433 \pm 0.008$ for B, and
$0.464 \pm 0.008$ for C, and $0.488 \pm 0.008$ for D, where we used an
average form $\overline{q^{2}}\equiv \int _{0}^{\cal T} d \tau q(\tau
)^{2} / {\cal T}$.  The expectation value of the mean squared radius
approaches to the ground-state value $0.5$ as the scale constant $\xi$
gets small and hence structures of the quantum fluctuations in fine
scale can be described, which is similar to the case of the
discretized approach.  
Then let us visualize degree of the quantum fluctuations by making a
histogram for $q$ with division of time range into sufficiently many
parts in each path and then combine $N=400$ histograms to obtain the
$q$-distribution $\rho _{q}$.  The result for the condition D is shown
in Fig.~\ref{fig:HO}(left) together with the squared wave function of
the ground state.  As one can see, our $q$-distributions in the
condition D behave consistently with the squared wave function.
Especially it is interesting that behavior of the quantum fluctuations
to large $q$ ($\sim \pm 3$) is very similar to the squared wave
function.  Furthermore, from the fluctuation one can evaluate the
correlation function defined as $C(\tau ) \equiv \Expect{\overline{q
    (\tau ) q(0)}} = \int _{0}^{\cal T} d \tau ^{\prime} q(\tau + \tau
^{\prime}) q(\tau ^{\prime}) / {\cal T}$.  The correlation function in
the condition D is shown in Fig.~\ref{fig:HO}(right) together with the
exact value for the grond state $= 0.5 \times \exp (- \tau)$.  As one
can see, the correlation function in condition D is quite similar to
the exact value up to $\tau \sim 2$ and the behavior is consistent 
even in larger $\tau$.

In the examination of our approach for the harmonic oscillator, we
have seen that our approach reproduces quantum behavior of the system
with more than about $95 \%$ accuracy by properly chosen scale
constants.  Especially, one can obtain higher accuracy by using
smaller scaling constant $\xi$.

Our approach can be applied to relativistic field theories.  It is
important that our approach has possibilities to become a
nonperturbative way to quantum field theories. To be specific, we
simulate $\UN{1}$ and $\SUN{2}$ gauge fields in four dimensions with a
periodic boundary condition (for the detail of the simulation
condition, see Ref.~\cite{Sekihara:2012ng}).  In this study we do not
include gauge fixing terms nor the Faddeev-Popov ghosts in the
Lagrangian densities.  Although the gauge invariance is maintained if
one takes into account all the possible paths for gauge fields, this
is not the case in our approach with finite $N_{\rm sum}$.
Nevertheless, the gauge group manifold is expected to be effectively
taken into account if one takes sufficiently large $N_{\rm sum}$ with
small $\xi$ in our approach.
One of the interesting results that at the saturation point
$\Expect{\overline{\cal L}_{\SUN{2}}} \approx 0.49 ~\xi^{-4}$ is
smaller than $3 \times \Expect{\overline{\cal L}_{\UN{1}}} \approx 3
\times 0.20 ~\xi^{-4}$ due to the self-interactions in $\SUN{2}$,
where $\overline{\cal L}$ is averaged Lagrangian density
[$\overline{\cal L}\equiv \int d^{4}x {\cal L}(x)/{\cal V}_{4}$ with
the four dimensional volume ${\cal V}_{4}$].  Furthermore, from the
Wilson loop~\cite{Wilson:1974sk}, we can evaluate a potential between
the (infinitely heavy) fundamental representation and its
antiparticle, which results in the Coulomb force and a confining
linear potential for the $\UN{1}$ and $\SUN{2}$ gauge fields,
respectively.

\section{Summary}

In summary, we have developed an approach to evaluation of the smooth
path integrals, in which paths are described by sum of smooth
functions with weight $\exp (-S)$ by appropriate methods.  In this
study we take an approximation that smooth fluctuations are described
only by the Gauss function.  The weighted smooth paths reproduce
properties of the ground-state harmonic oscillator in one dimension
with high accuracy by properly chosen width of the Gauss functions,
and the accuracy gets higher by using smaller width of the Gauss
functions, with which finer structure of the quantum fluctuations can
be described.  We have also evaluated quantum fluctuations of the
$\UN{1}$ and $\SUN{2}$ gauge fields in four dimensions and the Coulomb
force and a confining linear potential have been extracted from the
$\UN{1}$ and $\SUN{2}$ gauge fields, respectively.

\end{document}